\documentstyle[11pt,epsf,rotating,glas]{article}
\begin{document}

\begin{titlepage}{GLAS--PPE/96--02}{\today}


\title{Particle Physics in International Collaboration}

\vspace*{1.0cm}
\centerline{D.H. Saxon}
\centerline{\footnotesize Talk given at Kolloquiumstag ``125 Jahre
Teilchenphysik in Aachen", October 1995}

\end{titlepage}

\section{The first phase: 1947-1974}

        The modern era in Particle Physics began in 1947 in Bristol and
Manchester, with the discovery of the $\pi$-meson, of strange particles, and of
muon catalysed fusion \cite{ref1,ref2}. The first paper that describes the strong
interactions of a meson has only one author \cite{ref2}, but very quickly the
advantages of international collaboration came to the fore. Within a year a
American balloon (from Oppenheimer) was carrying a British emulsion (from
Bristol) into the stratosphere to look for cosmic ray interactions.

        Heisenberg's Uncertainty Principle gives the key to everything that
follows. The search for the ultimate constituents of matter leads to the
search for objects of zero size. (Causality and Relativity together dictate
that all finite size bodies must have the possibility of internal movement
and so must be composite.) In general then, one is looking for constituents
which are smaller than the composites that can be made from them, and the
Uncertainty Principle tells us that to access smaller distances one needs
higher energies. High energy particles are available in cosmic rays, but in
a totally random way so that major investment in detector technology would
be needed to get beyond what can be learned from the first observations. 

        Progress therefore required the construction of high energy
accelerators and that needed capital investment. At a UNESCO meeting in
Florence in June 1950, I.I. Rabi pointed out the ``urgency of creating
regional centres and laboratories in order to increase and make fruitful
the international collaboration of scientists in fields where the effort of
any one country is insufficient for the task." This meeting led eventually
to the setting up of CERN, the European Organisation for Nuclear Research,
initially with twelve member states, and now grown to nineteen. By June
1952 Geneva had been picked as the site for the laboratory, favoured over
Copenhagen, Paris and Arnhem. It is rumoured that the low costs of
operating in Switzerland were an important factor.

        By the 1960's a pattern had emerged of about ten ``state of the art"
accelerators at national and international laboratories around the world.
The most sophisticated experiments involved the use of bubble chambers.
Each chamber was an expensive capital facility capable of producing
exquisitely detailed pictures of the reactions of a beam of particles
incident on the liquid in the chamber (usually liquid hydrogen), held
momentarily in a superheated condition so that fine trails of bubbles
formed along the paths of charged particles. By observing the paths, one
could infer the motion of the particles, rather as one can observe the
passage of aircraft from the vapour trails in the sky. By varying the type
and the energy of the incident particle, a wide range of observations could
be made. With a production capacity of millions of photographs per year,
such a chamber could support a varied sequence of investigations, rather
like an astronomical observatory. 

        One took the film back home to one's institute, to scan and measure
with the assistance of skilled young ladies. Usually a group of
laboratories came together to make a common project. By sharing the
repetitious labour they could increase their measurement
capacity and build up a common library of data. They took the film first,
over a few days or weeks, and then shared the reels around the
collaboration. Problems of optical distortions and the like, which affected
the quality of the results, were different on film taken on different
dates.  Each site therefore took responsibility for certain reels of film,
measuring events of interest not only to themselves, but on processes whose
scientific explanation would be elucidated elsewhere. This usually took
more than one year, and technical analysis problems had to be solved as
they arose during this time. Sharing the development of software
tools resulted in both gains and losses in efficiency: some skill in
managing a collaborative effort is needed if such a venture is to be
coordinated around a set of Universities spread perhaps over two or three
countries.

        The relatively short time spent at the accelerator taking the film,
its transportability, and the fairly standardised nature of the
film-analysis technique encouraged collaborative work and allowed the
intellectual input of groups from far apart to be brought together to
improve the quality of the analysis. The sharing of responsibility for the
quality of the data, and the pooling of the data used in each analysis,
created a custom that the collaboration publishes its work as a whole.
Before one's work can be published, one must convince the collaborators
of its validity. There may be competing analyses on the same topics from
other groups within the collaboration, and these must reconciled before the
work can be made public. This internal refereeing has done more to raise
the quality of the field than any other single custom. It also laid the
foundations for the thousand-member collaborations of today. As always, the
external competition from other collaborations is vital, both as a safeguard
 and as
a spur.

        My own entry to the field, in 1966, was in search of a ``one-man"
experiment. I was afraid of the dangers of the big-science culture. In the
event I did complete a bubble chamber experiment substantially on my own,
with the advice and help of my supervisor and a sabbatical visitor, but
apart from the facilities in Oxford, where I was based, I used laboratories
in three countries - accelerators and bubble chambers at Rutherford (UK)
and Saclay (France) and measuring and computing facilities at Berkeley
(USA). Apart from an education in becoming street-wise during riots in two
countries, I mastered such topics as inventing tape formats that were
compatible between computers with 60-bit word length (in Berkeley) and
36-bit word length (in Oxford.) 8-bit bytes had yet to make their mark!

        Bubble chambers dominated Particle Physics Research in the 1960's.
But another style of work grew alongside it, using electronics and spark
chambers. This demanded a different culture. Teams of up to a dozen
scientists worked long hours for months at a time at the accelerator,
mastering technical problems to get a specialised detector ready to measure
accurately and with known efficiency a particular process, probably one not
accessible to bubble chambering because it demanded, say, measurement of
photons or neutrons or the use of a polarised target. I practised this
craft between 1974 and 1978. It teaches one how the quality of the detector
instrument dictates the range and usefulness of the studies that can be made.

        The style  of experimentation is driven by the science to be done.
Each generation of machines brings improves in energy or intensity that
demand new practices from the experimenters. The start-up of Fermilab (an
incremental process around 1972) marks a transitional phase. Experimental
teams were still small, but beginning to be challenged by the required
advances in technology. I myself went, in common with four other European
Research Associates, to work for Columbia - in New York for a year and then
at Fermilab, mastering such arts as cable tray and shielding block design!
I joined a Columbia-Fermilab experiment led by Leon Lederman, that was
looking for the $Z$ and $W$ bosons.

        We didn't know then how massive these bosons might be,
 but we planned to
produce them in proton-beryllium reactions at the highest possible energy
and intensity and to identify them by their decays to electrons and muons.
For the first time we used magnets, calorimeters and presamplers to
distinguish electrons and muons from pions that were ten thousand times
more common \cite{ref3}. The work on technique was innovative, and paved the way
for much of today's, but our effort was limited to building a spectrometer of
modest aperture, and was delayed by poor performance of the accelerator.
By April 1974, we had measured successfully inclusive electron
and muon production rates as a function of transverse momentum, and we were
working to build a second arm for the spectrometer, to detect particles
that decayed to a pair of leptons.

        We were too late. High energy physics was revolutionised in a
single day in November 1974, by the discovery of the $J/\psi$ particle
\cite{ref4}. It
was found in two separate ways, by hadronic production of lepton pairs,
just the method we had developed at Fermilab \cite{ref5}, and by the study of
electron-positron annihilation \cite{ref6}. The Fermilab group recovered, and
pushed on to discover the $\Upsilon$ particle \cite{ref7},
 but the success of the
electron-positron annihilation work changed not only our philosophy of
physics - within a few months the quark model became accepted as the norm
- but also the techniques.

\section{Colliding beam facilities: 1974 to the present.}

        ``$e^+e^-$ annihilations are rather poor producers of meson resonant
states and not likely to contribute much new knowledge of their
properties" \cite{ref8}.  In April 1974, one could say this without a qualm. By
November, nature had overturned it spectacularly, and for over twenty years
electron-positron colliders have dominated the field. The bosons produced -
the $\psi$ and $\Upsilon$ families and the $Z^0$ boson -
 have been studied with a precision
that outstrips virtually all else, and the current standard model is
dominated by our knowledge of their properties. 

        The SPEAR magnetic detector, which was used to make the $\psi$
discovery, is illustrated in figure 1. The SPEAR team
 could not use a bubble chamber
for colliding beams - the target, the duty cycle - everything was wrong.
But they attempted to build a ``4$\pi$" detector which copied its property of
seeing all the things which emerged from each reaction, whatever their
directions. No-one had attempted such an ambitious thing before using
electronic/spark chamber techniques. But this was what was needed to access
colliding beam physics, and to build and run it needed ``a new way to do
physics" (W Chinowsky, 1974) - the mega-team, in this case twenty-nine
people, was here to stay.

        By 1977, I knew that my polarised target experiments were coming to
an end at the NIMROD accelerator in England, and I needed to look elsewhere
for the future. I did one thing right, in deciding to work at PETRA, the
new high-energy electron-positron ring being constructed in DESY, Hamburg,
to run from late 1978. The lessons of the November Revolution had been well
learned, and this was a new step in complexity and in organisation. The
detector was full-solid angle. As at SPEAR, one took all reactions
simultaneously, with the same democracy as a bubble chamber, but with a new
collision to be inspected every 4 $\mu$s, rather than every two seconds or so.
This placed great importance on the event ``trigger". 

        One needed to reject unwanted collisions and select wanted
candidates at this prodigious rate, and needed to do this in a reliable and
understood way, with a known efficiency, known background, and known
biases. Electronic logic had developed to the point where complex
track-finding triggers could be built to work at this speed. But to make
this work in any sensible way, one had to be able to measure the
efficiency, to monitor it, and to make improvements. So one needed to
reconstruct the results from day one. The lead-time on event reconstruction
 fell from 18
months to one day, as the trigger experts clamoured for feedback on their
performance. The higher energy of the collisions forced every part of the
detector to be bigger than at SPEAR. 

        How to achieve all this? A collaboration of powerful institutes was
needed, each with expert teams specialising in certain aspects of the task.
In the case of the TASSO collaboration, we began with 87 scientists from
nine institutes in three countries: the number of institutes and countries
grew a little later. It was a marvellous experience - the thrill of
producing new science in all directions, the stimulus of working with
enthusiasts from so many places, and the urgency of the competition inside
and outside the collaboration, (a spur both to speed and to accuracy,) all
combined to raise the level of our work.

        A particle physics experiment is a long undertaking, though not
perhaps as long as the time scales involved in the design, construction and
exploitation of scientific satellites. It starts with a scientific idea,
perhaps amongst a group of friends. The development of the detector
technique can take over five years - for the quality of the instrument is
paramount to the range of the work that can be done. If one takes HERA as
an example, meetings on possible electron-proton colliders were held as
early as 1975. The subject acquired real momentum (as viewed by this
experimenter) in Genoa in 1984, and the first results came in 1992. The
full exploitation of the facility will take over ten years.
                                       
Particle Physics experiments are like astronomical observatories, in that
they permit a range of studies. But by contrast with observatories, each
study addresses the entire data sample, accumulated perhaps over several
years. Wanted data are selected by triggering and by appropriate cuts in
analysis. So different subsystems, designed to measure different aspects of
each event, must cohabit without degradation of each others' performance.
Thus semiconductor `vertex detectors' which measure track positions close
to the production point must have minimal material to reduce both multiple
scattering and photon conversion which would degrade the track momentum and
multiplicity measurement.

        The more sophisticated results are often obtained by combining
results from different detector components. An example from the LEP
collider at CERN, producing a $Z^0$ boson which decays to a
quark-antiquark pair, is shown in figure 2. This shows the production of a
$B_s$ meson, consisting of a $(b \overline{s})$ quark pair. The $B_s$
 decays after travelling
about 3 mm, to $\psi^{\prime} \phi$. The $\psi^{\prime}$ then decays to  $\mu^+ \mu^-$,
 and the $\phi$ to $K^+K^-$. The
main tracking chamber (`time projection chamber') was used to measure the
particle momenta, and to identify $K$-mesons by their ionisation loss, the
vertex detector to separate clearly the production and decay of the $B_s$, the
electromagnetic calorimeter identified the electron, and the penetration of
the particles through the hadron calorimeter into the muon chambers
identifies the $\mu^+$ and the $\mu^-$.

        The ALEPH collaboration, which worked together to construct and
exploit this elegant detector, started work in 1982 and now comprises over
400 physicists from 32 institutes in 10 countries.\cite{refA}
 Not all these countries
are member states of CERN. People outside the field often ask how such a
body is managed. It is done using no documents of greater legal strength
than a memorandum of understanding - a document which binds the funding
agencies of each country to best endeavours, but is not something on which
you could sue. The collaborations operate within a formal structure devised
by themselves under the avuncular gaze of the CERN management. In the end,
it works because the scientists themselves want it to work. The sanction
for failure to deliver results, for poor performance or inadequate maintenance of
 equipment, or for
unethical behaviour is loss of reputation. 

        The world of Particle Physics is in one sense small. The fear of
being drummed out of the club is a powerful sanction, but the very
smallness of this world makes it in another sense large. It provided some
porosity in the iron curtain long before its rust began to be evident
elsewhere, and continues to provide a mechanism for cooperation between
both well resourced and poorly resourced countries. Groups from afar that
bring little in the way of Swiss Francs can contribute as equal
collaborators if they contribute intellectual capital that their partners
value.

\section{New projects: 1995 on}

        Investigative science must progress or die. For Experimental
Particle Physics this is especially true. Governments will tolerate
continuing to pay our bills only if they see continued vigour and advance.
Decisions to build new accelerators must be driven by theoretical insights
that point the way forward. The need for high energies to study neutrinos
impelled the construction of the SPS. LEP was impelled by precisely
testable predictions of electroweak theory, HERA by questions on the
structure of matter and its interactions, and the Large Hadron Collider by
theoretical insight (the Higgs mechanism) on the origin of mass and hints
of higher symmetries.

        The financial climate has become less congenial. In the 1970's the
CERN SPS was built on a dramatically increased annual budget. In the 1980's
LEP was built within a constant annual budget. In 1994 CERN Council
insisted that the Large Hadron Collider be built within a shrinking budget,
with special contributions from the two countries in which the LHC will sit
(across the Franco-Swiss border) and with substantial contributions to help
and accelerate the construction of the collider from non-member states
who wish to share in its exploitation.

        One must therefore be ruthless in moving on when a current venture
has fulfilled its potential. LEP ran from 1989-1995 as a $Z^0$ factory and
made measurements of unparalleled impact over a range of topics. The
precision that has been achieved is legendary and has transformed the style
of investigation. The beauty of the measurements that established the
sensitivity of the measurement of the $Z^0$ mass to the tidal distortion of
the earth's crust by the moon has subsequently revealed effects due to
rainfall, to the height of the water in Lake Geneva, and to the daily
passage of trains between Paris and Geneva! The precision of the detectors
allowed the identification of processes such as that shown in figure 2,
 but did not provide sufficient statistical power or sufficient
energy to pursue the now-visible goals in fermion and boson physics. 

        To pursue these goals further two different machines are needed.
LEP is being upgraded in stages from 1995 to 1999 to study the $W±$-boson and
to search for new phenomena that are hinted at in the Higgs and
supersymmetry sectors. The LEP2 machine (as it is called) is a bargain. The
LEP accelerator and all its infrastructure is reused with the addition of
superlative new superconducting accelerating cavities. (The quality of
these can be expressed numerically. An organ playing concert-pitch A and
built to the same standard would reverberate for a year after the note was
sounded - an acoustic that puts any cathedral in the shade.) The nineteen
member states that together pay for CERN can take pride in the centre of
excellence they have created that can make these. 

        The detectors already built for $Z^0$ studies needed only modest
upgrading to handle the higher energies and rates.  But the upgrading is
also of the highest quality, and its realisation is an example of the way
international collaboration works. Consider the ALEPH vertex detector
upgrade (an example with which I am familiar.) Compared to the vertex
detector used for $Z^0$ work, this was devised to extend the acceptance so as
to make most efficient use of the accelerator, to reduce the material
intercepting outgoing particles whilst providing additional measurements,
and to withstand higher radiation doses. The upgraded system has Italian
detector elements,  French mechanics and British electronics. To minimise
material whilst maximising thermal conduction to carry away heat beryllia
substrates and diamond-loaded glue were used.

        But the trail of quark-physics found by LEP whilst studying the $Z^0$
cannot be pursued to the needed statistical precision at LEP.
 This needs a new accelerator optimised for high
intensity rather than high energy. The energy (10 GeV mass rather than 90
GeV as at LEP) is optimised to produce $B$-mesons in a particular controlled
way. The topic addressed is vital for our understanding of the Universe. It
offers a clue as to how a Big Bang dominated by radiation (`let there be
light') could have evolved into one of matter such as we live in now. And
it demands a purpose-built accelerator.

        A number of European groups are moving to new facilities, most to
California to the Babar experiment being constructed at Stanford.
Competition with another facility under construction in Japan will
guarantee the vitality and quality in the work. Cooperation across the
Atlantic is well-developed amongst scientists, though not, it appears, so
easily addressed by certain governments whose modes of authorisation and
funding are rather different. Cooperation between Europe and Japan has been
highly successful (for example at PETRA and at LEP), but the scale has so
far been more limited. It is good to see the level of cooperation now begun
between Japan and CERN as an augur for the future.

\section{The HERA model and the Large Hadron Collider.}

        The construction of HERA - the Hadron Elektron Ring Anlage - at
DESY between 1984 and 1992 provided a working example of the way forward
for international and inter-regional collaboration. Germany announced its
firm intention to build the new electron-proton collider and invited other
countries who wished to do so to join in the construction. In the event a
number did by constructing elements in their own countries that formed part
of the machine, most notably the construction of 3 km of superconducting
magnets by Italy. Other substantial components were contributed by Canada,
France, Israel, the Netherlands and the USA. Major expert assistance was
supplied by Poland and China, and Czechoslovakia: the DDR and the United
Kingdom also supplied expertise. Countries who did not contribute in a
major way to building the machine could contribute to the experiments, but
subsequently pay certain running costs.

        CERN is following a similar model for the Large Hadron Collider.
The full project for an energy of 14 TeV was approved by CERN Council in
December 1994. Based on the member-state funding foreseen at that time it
will operate at reduced energy from 2004, and at full energy from 2008.
Non-member states who wish to use the accelerator are invited to contribute
in cash or kind in a way that will bring this timetable forward. The
invitation is open, but there is an expectation this time that such
countries will indeed contribute to the machine.

        Japan responded with an immediate gift of 5BYen,
 perhaps with more to
follow, and agreements have been reached, or are well advanced, with
Canada, Israel, India and Russia. Negotiations with the USA are underway,
with a view to reaching an agreement before the progress of the LHC project
is scheduled for review in 1997. The somewhat harsher financial climate of
the 1990's is part of the explanation of this stiffening of attitude,
together with the perception that it is some years since there was a broad
balance in the two-way 
flow of scientists across the Atlantic, at least within the
field. The USA continues to offer entry to Babar on the basis of
contributing to detector and associated running costs but not to the
capital cost of the accelerator and European scientists are certainly keen
to work there, though by no means as many as the Americans who want to use
the LHC.

        But HERA provides a pointer to the future in another and an
important way. Table 1 illustrates this. It shows the time intervals at
which beam-beam crossings occur at different storage rings, and the
expected number of interactions per crossing (HeraB is a high-rate
$b$-physics experiment being constructed at HERA to run in 1997). Every
crossing must be interrogated to see if it contains a useful interaction.
One can record only a few events per second. HERA puts us for the first
time into a situation where the interval between crossings that must be
interrogated is less than the time it takes to make a decision, and in fact
less than the time it takes to read the raw data off the detector into a
buffer.
\begin{table}[htb]
\begin{center}
\begin{tabular}{l c c c}\hline
Facility & Year & Crossing Interval & Events/crossing \\ \hline
                PETRA    &      1978       &    3.6 ms       &          rare \\
                LEP      &      1989       &    22 ms        &          rare \\
                HERA     &      1992       &    96 ns        &          rare \\
                HeraB    &      1997       &    96 ns        &          4 \\
                LHC      &      2004/8     &    25 ns        &          25 \\ \\ \hline
\end{tabular}
\end{center}
\caption{ Rates of potential and actual interactions at various storage rings}
\end{table}

        Real-time event selection and processing is therefore needed.
Unwanted crossings must be rejected, and wanted ones identified. These
 form only one
part in $10^4$ or $10^7$ of the total number of candidate crossings,
 for a range of relatively
common or rare processes. The ``trigger" needs high and known efficiency,
and the ability to develop to counteract backgrounds or to accept processes
not considered at the original design phase. One sits on a learning curve,
first from HERA to HeraB to LHC on the use of ``pipelined logic" and
secondly in the early part of the running of the experiment, on developing
triggers of increasing sophistication.

        Figure 3 indicates the logic structure proposed for the ATLAS
experiment for the large hadron collider\cite{ref9}.
 The data flows are tremendous.
The event sizes are such that the front-end data flow exceeds that of
present day television sets in the whole of Europe. Successive
trigger levels have more time to think, and allow increasingly complex
logical decisions to be made on the data. Level 3 involves full event
reconstruction. 

        Figure 4 shows the `learning curve' for the ZEUS experiment. Data
flow restrictions force the first level trigger rate (FLT) to be below a
few hundred Hz and the output to tape (TLT) to be below about 10Hz. At the
start of running in 1992 the rates for all trigger levels were very similar
and independent of the (very low) luminosity (and so were
background-dominated). By 1994 the techniques had become much more
sophisticated. Rates are far below linear extrapolations with intensity
from 1992, whilst a broader range of physics processes is addressed. The
lessons for LHC start-up, when a variety of presently foreseen or
unforeseen processes will need to be studied that 
are excluded by simply selecting on the highest transverse energy, are
clear. The HeraB experiment will provide the next test of these triggering
ideas\cite{ref10}.
 From 1997 four events every 96 ns will have to be tested to see
if they are the 1 in $10^5$ in which $b$-quarks are produced.

        To handle the energy, the rate, and the complexity of the events
the LHC experiments will also be large and complex. Figure 5 illustrates
the structure of the CMS experiment \cite{ref11}, and figure 6 shows a typical
simulation of an event reconstructed in the ATLAS inner tracker. The
technical demands imposed by the LHC on detector technology have stimulated
work on detector Research and Development across the whole of the CERN user
community, whether on radiation hard semiconductor pixel detectors, high
resolution electromagnetic calorimetry, fibre-optic data links,
object-oriented programming, cooling and alignment or parallel computing
\cite{ref12}.
 The `technology-foresight' and industrial links provide a valuable
spin-off from the field.

\section{The outlook.}

        The LHC marks perhaps a transitional phase into world-wide
collaboration. Some
30\% of the signatories to the two proton-proton experiments are from
 non-member states. There are 73 nationalities working in
institutes in 54 countries. Of these the USA accounts for 507 people.
Russia 362, Canada 60, India 42, Japan 38, China 34, Georgia 31, Belarus
28, Israel 22 and Bulgaria 22. All have different local resources,
histories and possibilities. No single structure for cooperation agreements
will suit all these. The political dimensions have developed both within
countries and in supranational structures.

        The European Committee for Future Accelerators (ECFA) has provided
an important forum for developing proposals for new machines, whether at
CERN or elsewhere. It has a brand-new counterpart in Asia (ACFA). The
International Committee for Future accelerators (ICFA) provides a mechanism
for cooperation between the directors of major accelerators world-wide.
ICFA has blessed the formalisation of collaboration on the development of
accelerator techniques for future linear electron-positron colliders: 23
laboratories in Asia, Russia, USA and CERN member states are involved.
Comparison of LEP and the TeVatron shows that the reach in energy of the
proton machine is higher, but that electron machines can grasp specific
topics with great precision. The top-quark discovery \cite{ref13} creates
complementary virtues in proton-proton and electron-positron machines.To
reach energies of 250 GeV or more in a reasonable length (say 20 km)
accelerating fields over 10 MV/m are needed, and at a cost we can afford.
Once this technical work has succeeded, it is more than likely that the
strong support of laboratory directors in several major countries will be
needed if such a project is to gain the required  breadth of support across
governments.

        The alternatives to cooperation are atrophy, or at best external
direction and delay. The OECD Megascience Forum has looked at Particle
Physics, and has flirted with the idea of baskets of projects across a
range of fields. (Reference \cite{ref14}
 provides some background.) This might be
attractive at a time of rapidly increasing wealth, but can run into the
sand at times of stringency. The International Union of Pure and Applied
Physics has argued strongly that science should not be the exclusive
preserve of the wealthiest nations.

        One of the products of Particle Physics most valued by governments
is highly trained manpower. In the United Kingdom it is viewed in
the following way. Students commence Ph D training at age 22 after four
years of undergraduate training. They are financed for three years, and
expected to complete their work promptly.
 Indeed the Universities are subject to penalties if
students take more than four years. Compared with countries where the Ph D
takes much longer, there are plenty of good applicants and when they finish
their Ph D's their training is valued by employers. Working in a
multi-national collaboration gives them leading-edge technical skills,
project management skills, communication skills, team-working skills and 
self reliance. With Physics in its present stage of evolution, we have an
excellent case for the number of research students to be increased.

        Throughout the changes in Particle Physics since 1947, our
techniques have always pursued the most advanced technology.
 We must conclude that international collaboration
has proved a robust tool. It increases both the strength and the quality of
the activity. Particle Physics has maintained its vigour, its scientific
excellence, and its educational value. It has withstood rigorous scrutiny. We
have every reason for confidence in the future, provided that we understand
the changing environment in which our work takes place.

\begin{figure}[htb]
\hspace{3.0cm}
\epsfig{file=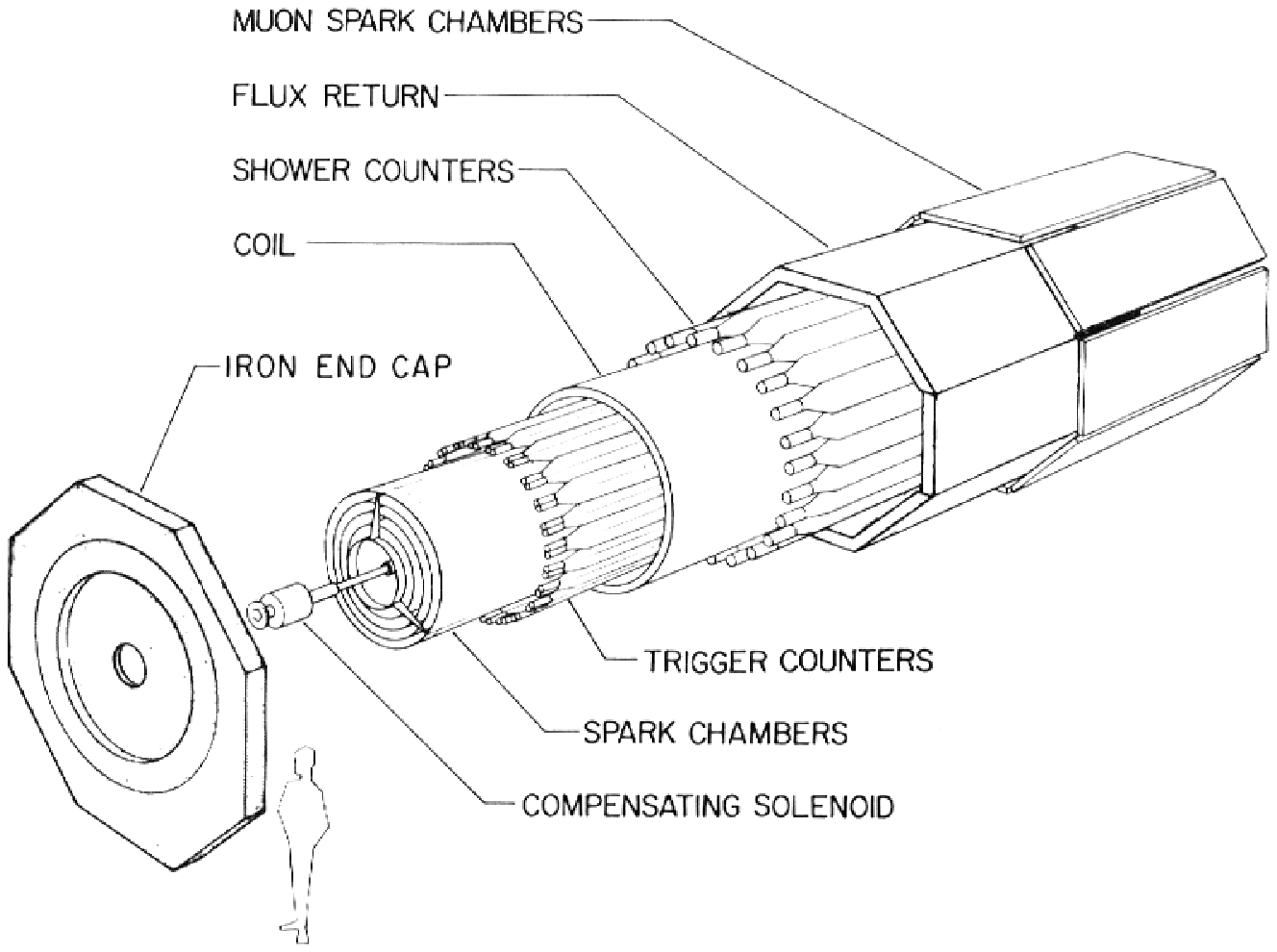,height=8.0cm}
\epsfig{file=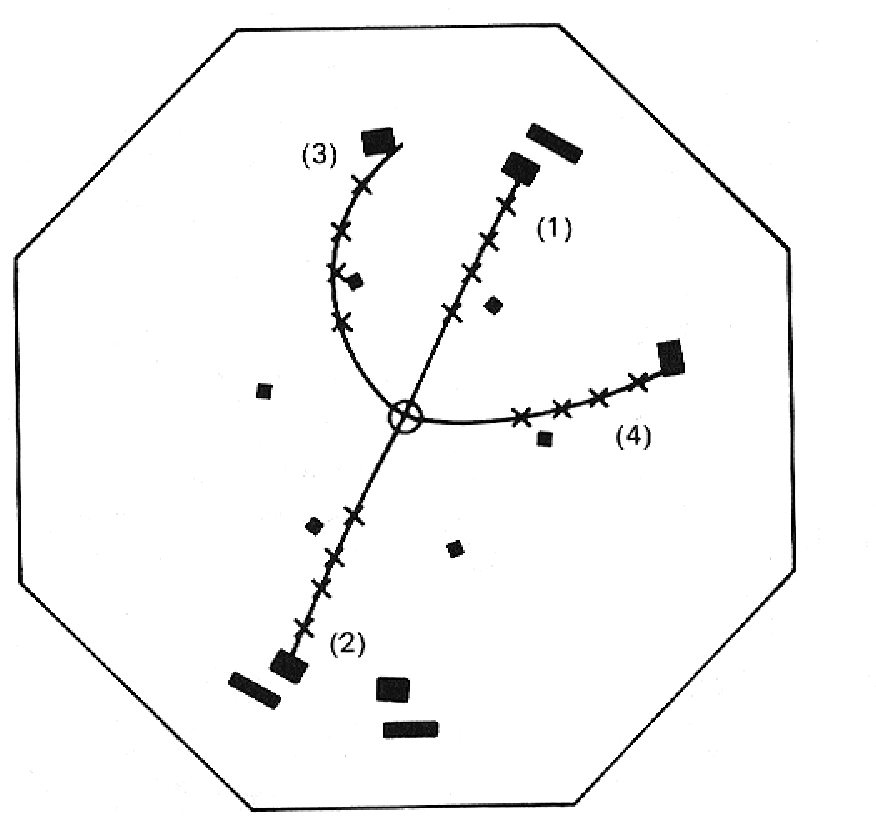,height=8.0cm}
\caption{ SPEAR magnetic detector. (a) schematic layout (b) a event of the type
$e^+e^- \rightarrow \psi^{\prime}  \rightarrow  \psi \pi^+\pi^- $ ; 
$\psi \rightarrow
\mu^+\mu^-$. From the pattern of tracks, this was
jokingly called the `auto-signature' mode of the $\psi$.}
\end{figure}
\newpage
             
\begin{figure}[htb]
\hspace{3.0cm}
\begin{turn}{-90}
\epsfig{file=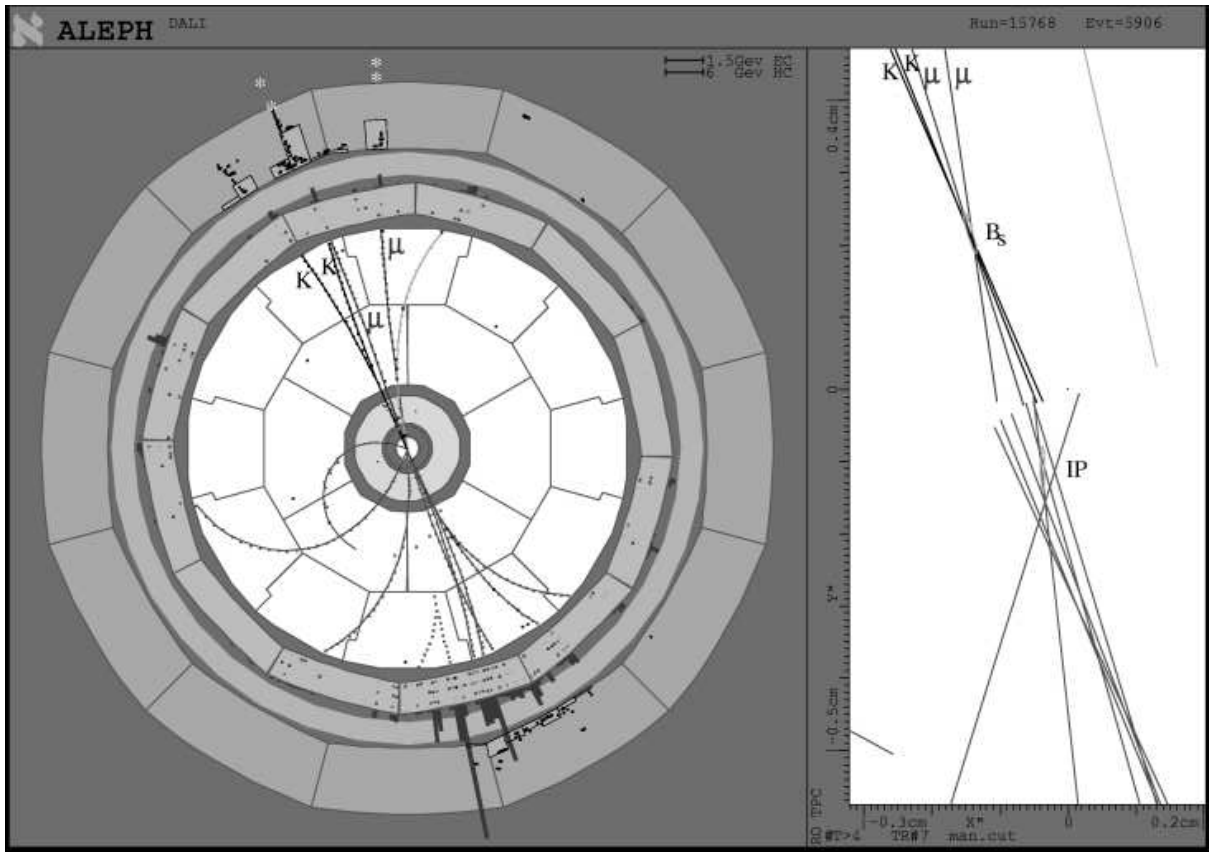,height=16cm}
\end{turn}
\caption{  An event reconstructed in the ALEPH detector - main view is of the
event seen in the Time Projection Chamber, with the electromagnetic
calorimeter, hadron calorimeter and muon detectors in successive layers
around it. The vertex region, as reconstructed using
the vertex detector is shown on the right, with a 3mm flight path from
the interaction point to the $B_s$ decay vertex.
 See text for details of the event.}
\end{figure}
\newpage
                                                       
\begin{figure}[thb]
\centerline{
\epsfig{file=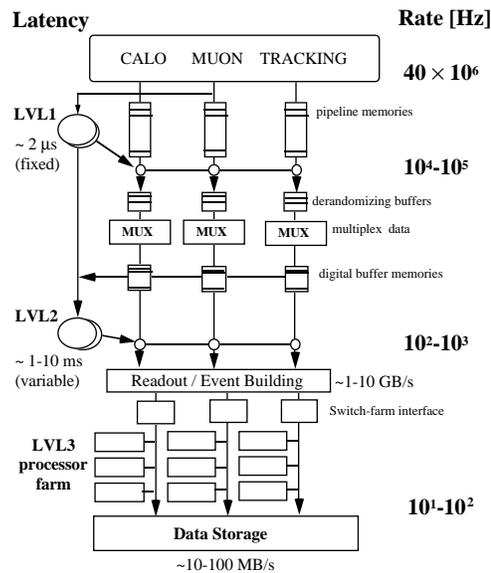,height=8cm}
}
\caption{ Trigger and Data Acquisition structure of the ATLAS experiment
being developed for the LHC}
\end{figure}
                           
\begin{figure}[htb]
\centerline{
\epsfig{file=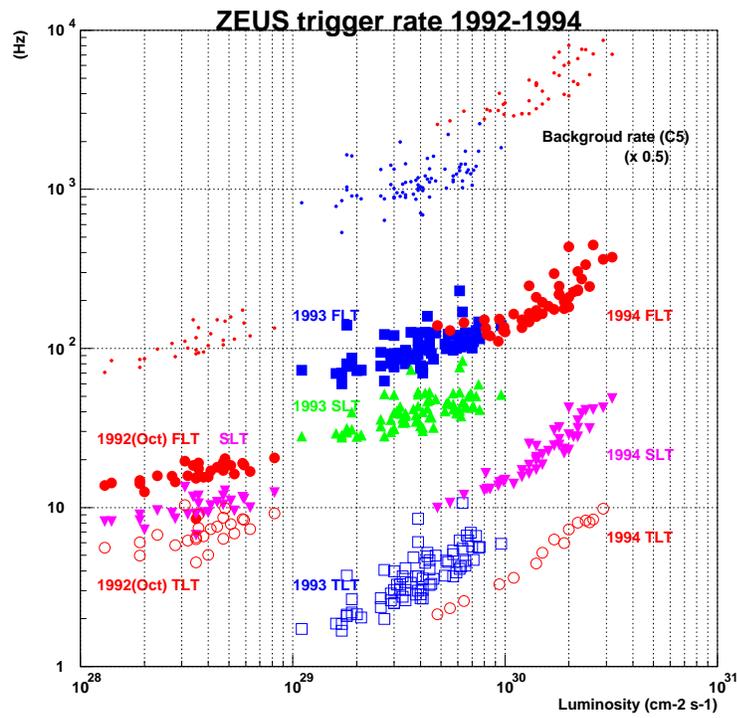,height=10cm,bbllx=10,bblly=150,bburx=565,bbury=675}
}
\caption{ ZEUS (HERA) trigger rates as a function of luminosity. FLT - first
level trigger, SLT - second level trigger, TLT = third level, output to
tape.   }
\end{figure}
\newpage
        
\begin{figure}[thb]
\centerline{
\epsfig{file=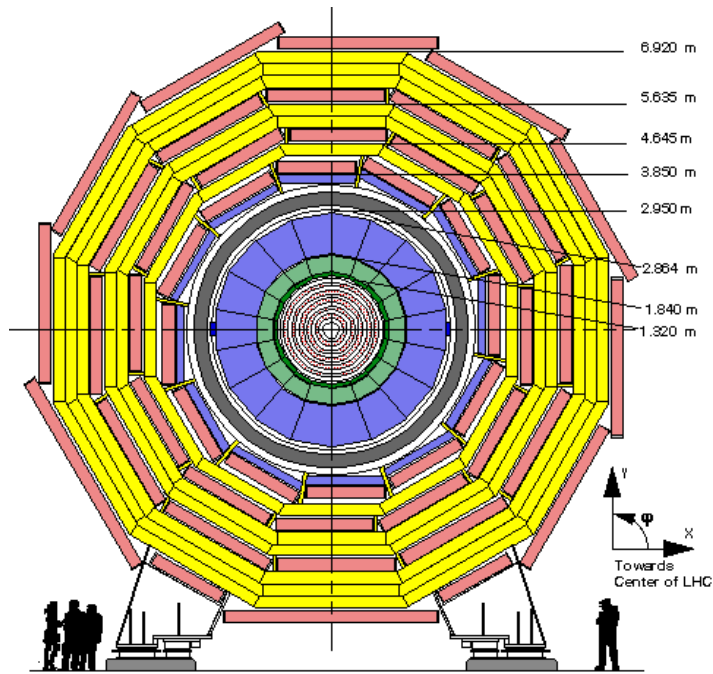,height=10cm,bbllx=30,bblly=390,bburx=410,bbury=750}
}
\caption{ Layout of the CMS detector as proposed for the LHC}
\end{figure}
                                                           
\begin{figure}[htb]
\epsfig{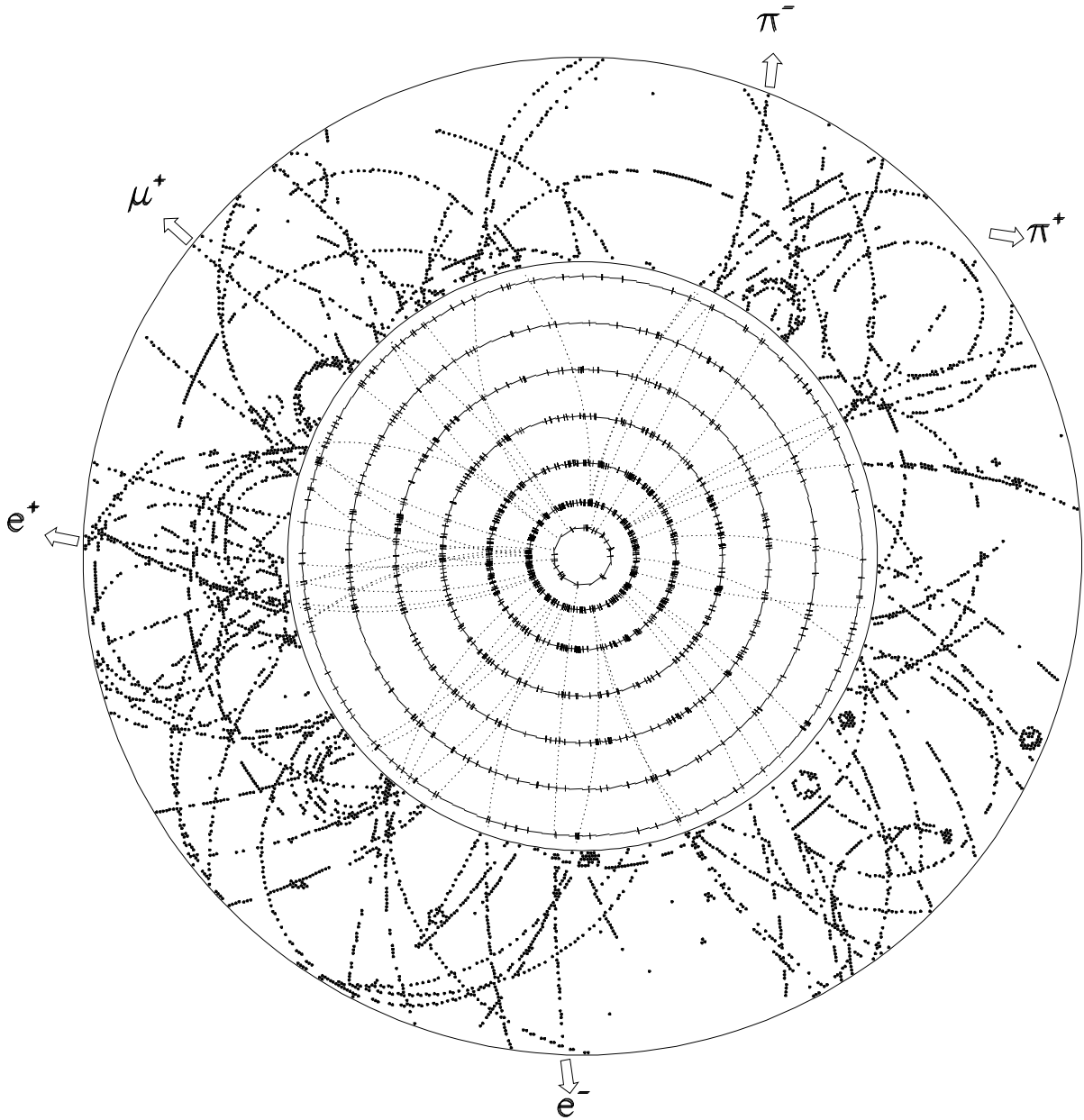}
\caption{ A simulated event pictured in the central part of the ATLAS inner
detector. The inner layers are silicon tracking detectors with 80 $\mu$m
granularity. The outer layers are 4 mm diameter proportional drift tubes
(`straws') The event shown is at a luminosity of $5.10^{33} {\rm cm^{-2}s^{-1}
}$, and includes the production of a $B^0_d$ meson decaying to $J/\psi K^0_s$,
with $J/\psi \rightarrow e^+e^-, K^0_s \rightarrow \pi^+\pi^-$.}
\end{figure}

\end{document}